\documentclass[%
 reprint,
 amsmath,amssymb,
 aps,
 prl,
 floatfix,
 superscriptaddress,
]{revtex4-1}

\usepackage{graphicx}% Include figure files
\usepackage{dcolumn}% Align table columns on decimal point
\usepackage{bm}% bold math
\usepackage{color}

\begin{document}

%%%%%%%%%% SHORTHAND LATEX DEFINITIONS %%%%%%%%%%

%%%%%%%%%%%%  MCGILL DEFINITIONS  %%%%%%%%%%%%%
\newcommand{\ud}{\mathrm{d}}
\newcommand{\rd}{\partial}
\newcommand{\ui}{\mathrm{i}}
\newcommand{\RK}{\mathrm{\scriptscriptstyle{RK}}}
\newcommand{\h}{\mathrm{h}}
\newcommand{\zerov}{\mbox{\boldmath$0$}}
\newcommand{\alphav}{\mbox{\boldmath$\alpha$}}
\newcommand{\piv}{\mbox{\boldmath$\pi$}}
\newcommand{\sigmav}{\mbox{\boldmath$\sigma$}}
\newcommand{\Vv}{\mbox{\boldmath$V$}}
\newcommand{\pv}{\mbox{\boldmath$p$}}
\newcommand{\Dv}{\mbox{\boldmath$D$}}
\newcommand{\Av}{\mbox{\boldmath$A$}}
\newcommand{\Ev}{\mbox{\boldmath$E$}}
\newcommand{\Bv}{\mbox{\boldmath$B$}}
\newcommand{\Lv}{\mbox{\boldmath$L$}}
\newcommand{\ev}{\mbox{\boldmath$e$}}
\newcommand{\Mv}{\mbox{\boldmath$M$}}
\newcommand{\Tv}{\mbox{\boldmath$T$}}
\newcommand{\tav}{\mbox{\boldmath$\tau$}}
\newcommand{\xv}{\mbox{\boldmath$x$}}
\newcommand{\Sv}{\mbox{\boldmath$S$}}
\newcommand{\hv}{\mbox{\boldmath$h$}}
\newcommand{\Iv}{\mbox{\boldmath$I$}}
\newcommand{\rv}{\mbox{\boldmath$r$}}
\newcommand{\kv}{\mbox{\boldmath$k$}}
\newcommand{\av}{\mbox{\boldmath$a$}}
\newcommand{\bv}{\mbox{\boldmath$b$}}
\newcommand{\cv}{\mbox{\boldmath$c$}}
\newcommand{\nablav}{\mbox{\boldmath$\nabla$}}
\newcommand{\sech}{\mathrm{sech}}
\newcommand{\so}{\mathcal{L}}

\newcommand{\avg}[1]{\left\langle #1 \right \rangle}
\newcommand{\unit}[1]{\,[\mathrm{#1}]}
\newcommand{\abs}[1]{\left|#1\right|}
\newcommand{\tm}[1]{\textrm{#1}}
\newcommand{\el}[2]{^{#1}\mathrm{#2}}
\newcommand{\mm}[1]{\mathrm{#1}}
\newcommand{\drv}[2]{\frac{\ud#1}{\ud#2}}
\newcommand{\id}[1]{\mathrm{d} #1\,}
\newcommand{\idd}[2]{\mathrm{d}^{#1}\!#2\,}
\newcommand{\bra}[1]{\langle #1|}
\newcommand{\ket}[1]{|#1\rangle}
\newcommand{\braket}[2]{\langle #1|#2\rangle}
\newcommand{\ketbra}[2]{\ket{#1}\!\bra{#2}}
\newcommand{\kdim}[2]{\ket{#1,\,#2}}
\newcommand{\bdim}[2]{\bra{#1,\,#2}}
\newcommand{\com}[2]{\left[#1,#2\right]}
\newcommand{\re}[1]{\mathrm{Re}\left[#1\right]}
\newcommand{\im}[1]{\mathrm{Im}\left[#1\right]}

\def \uc{\mathrm{c}}
\def \ue{\mathrm{e}}
\def \ud{\mathrm{d}}
\def \uo{\mathrm{o}}
\def \uq{\mathrm{q}}
\def \uI{\mathrm{I}}
\def \uR{\mathrm{R}}

\def \sx{\hat{\sigma}_x}
\def \sy{\hat{\sigma}_y}
\def \sz{\hat{\sigma}_z}
\def \sp{\hat{\sigma}_+}
\def \sm{\hat{\sigma}_-}

\def \ha{\hat{a}}
\def \hd{\hat{d}}
\def \hH{\widehat{H}}
\def \hI{\widehat{I}}
\def \hO{\widehat{O}}
\def \hQ{\widehat{Q}}
\def \hR{\widehat{R}}
\def \hS{\widehat{S}}
\def \hU{\widehat{U}}
\def \hV{\widehat{V}}
\def \hW{\widehat{W}}

\newcommand{\degr}{^{\circ}}
\newcommand{\epsIn}{\epsilon_\text{in}}
\newcommand{\epsOut}{\epsilon_\text{out}}
\newcommand{\GammaMeas}{\Gamma_\text{meas}}
\newcommand{\GammaMeasVac}{\Gamma_\text{meas,vac}}
\newcommand{\GammaPhi}{\Gamma_{\varphi}}
\newcommand{\GammaPhiVac}{\Gamma_{\varphi,\text{vac}}}
\newcommand{\OmegaR}{\Omega_\text{R}}
\newcommand{\sigmaX}{\hat{\sigma}_x}
\newcommand{\sigmaY}{\hat{\sigma}_y}
\newcommand{\sigmaZ}{\hat{\sigma}_z}
\newcommand{\omegaC}{\omega_\text{c}}
\newcommand{\omegaQ}{\omega_\text{q}}
\newcommand{\abar}{\bar{a}_0}
\newcommand{\tauPhi}{\tau_{\phi}}
\newcommand{\tauMeas}{\tau_\text{meas}}
\newcommand{\Tint}{T_\text{int}}
\newcommand{\TRabi}{T_\text{Rabi}}
\newcommand{\GSQZ}{G_\text{SQZ}}
\newcommand{\FigOne}{Fig.~\ref{intro_fig}}
\newcommand{\FigTwo}{Fig.~\ref{dephaseFig}}
\newcommand{\FigThree}{Fig.~\ref{measFig}}
\newcommand{\FigFour}{Fig.~\ref{etaFig} }

\newcommand{\QNLaffil}{\affiliation{Quantum Nanoelectronics Laboratory, Department of Physics, University of California, Berkeley, California, USA}}
\newcommand{\CQCSaffil}{\affiliation{Center for Quantum Coherent Science, University of California, Berkeley, California, USA}}
\newcommand{\McGaffil}{\affiliation{Department of Physics, McGill University, Montr\'{e}al, Qu\'{e}bec, Canada}}
\newcommand{\IMEaffil}{\affiliation{Institute for Molecular Engineering, University of Chicago, Chicago, Illinois, USA}}

%%%%%%%%%% HEADER STUFF %%%%%%%%%%

%\preprint{APS/123-QED}

\title{Stroboscopic qubit measurement with squeezed illumination}

\author{A. Eddins}
\QNLaffil
\CQCSaffil
\author{S. Schreppler}
\QNLaffil
\CQCSaffil
\author{D. M. Toyli}
\QNLaffil
\CQCSaffil
\author{L. S. Martin}
\QNLaffil
\CQCSaffil
\author{S. Hacohen-Gourgy}
\QNLaffil
\CQCSaffil
\author{L. C. G. Govia}
\McGaffil
\IMEaffil
\author{H. Ribeiro}
\McGaffil
\author{A. A. Clerk}
\McGaffil
\IMEaffil
\author{I. Siddiqi}
\QNLaffil
\CQCSaffil

\date{\today}

\begin{abstract}
Microwave squeezing represents the ultimate sensitivity frontier for superconducting qubit measurement. However, observation of enhancement has remained elusive, in part because integration with conventional dispersive readout pollutes the signal channel with antisqueezed vacuum. Here we induce a stroboscopic light-matter coupling with superior squeezing compatibility, and observe an increase in the room-temperature signal-to-noise ratio of 24\%. Squeezing the orthogonal phase controls measurement backaction, slowing dephasing by a factor of 1.8. This protocol enables the practical use of microwave squeezing for qubit state measurement.

\end{abstract}

\maketitle

%%%%%%%%%% INTRODUCTION %%%%%%%%%%
Electromagnetic quadrature squeezing is the reduction below vacuum noise of fluctuations in, for example, either the $\sin(\omega t)$ or $\cos(\omega t)$ component of the electric field. Besides being of fundamental interest as nonclassical states of light, squeezed states can enable faster measurements in cases where field intensity is limited, utilizing multi-particle quantum correlations to encode more information per photon, which manifests as a reduction in noise. Optimally applied squeezing can result in Heisenberg-limited scaling, wherein the signal-to-noise ratio (SNR) scales as the number of measurement photons instead of as the square root. The development of squeezing at optical frequencies has a long history \cite{Andersen201630Generation} leading to a range of recent applications such as gravitational wave detection \cite{Abadie2011ALimit, Aasi2013EnhancedLight}. Squeezing of microwave-frequency fields using superconducting amplifiers \cite{Yurke1988ObservationAmplifier,Castellanos-Beltran2008AmplificationMetamaterial,Eichler2014Quantum-LimitedResonators} has surged as a topic of interest in the modern contexts of circuit quantum electrodynamics and dark matter detection \cite{ZhengAcceleratingTechnology} given the ability to couple squeezed fields to low-dimensional quantum systems such as superconducting qubits \cite{Murch2013ReductionVacuum.,Toyli2016ResonanceVacuum,Kono2017NonclassicalDrive}, optomechanical circuits \cite{Clark2016ObservationOscillator,Clark2017SidebandLight}, or spin ensembles \cite{BienfaitMagneticMicrowaves}. 

Despite interest in using squeezed microwaves for superconducting qubit measurement, experimental realization has remained elusive. A major challenge is that the dispersive coupling central to standard readout techniques rotates squeezing out of, and antisqueezing into, the signal quadrature, limiting SNR improvement outside of certain restrictive parameter regimes \cite{Barzanjeh2014DispersiveAmplifiers}. Proposals to fully exploit squeezing for qubit measurement have been suggested, however with the need for more complex circuit architectures involving multiple readout modes \cite{Didier2015HeisenbergLight,Govia2017EnhancedSuppression} or a fundamentally different longitudinal qubit-cavity coupling \cite{Didier2015FastInteraction}. In this Letter, we employ a stroboscopic longitudinal coupling \cite{Hacohen-Gourgy2016QuantumObservables} compatible with large amounts of squeezing and standard qubit designs to harness input squeezing for qubit measurement. Stroboscopic techniques have been intensely studied in ``backaction evading'' optomechanical systems \cite{Hertzberg2010Back-action-evadingMotion,Suh2014MechanicallyField,Lecocq2015QuantumObject}, but have not been combined with injected squeezing.

Stroboscopic measurement occurs within a rotating frame in which the terms which limit the benefit of squeezing for conventional dispersive readout are suppressed. The scheme consists of a qubit Rabi-oscillating at frequency $\OmegaR$ coupled to a cavity driven by sideband tones at $\omegaC \pm \OmegaR$. The interaction-picture Hamiltonian is $\hat{H}_\text{I} = \chi\sigmaZ a^\dagger a + \frac{\OmegaR}{2}\sigmaX + \hat{H}_\text{sb}$, where $\hat{H}_\text{sb}$ describes the sideband drives. Decomposing the resulting cavity field into its classical and quantum parts, $\hat{a}\rightarrow 2\abar \cos(\OmegaR t) + \hat{d}$, and transforming to the Rabi-driven qubit frame yields
\begin{equation}\label{eqn:HRabi}
\hat{H}_\text{R} = \chi\abar\sigmaZ^\text{R}(\hat{d}+\hat{d}^\dagger) + (e^{i\OmegaR t}\hat{A}+e^{i2\OmegaR t}\hat{B}+\text{H.c.}).
\end{equation}
The first term of Eq.~\eqref{eqn:HRabi} describes a resonant longitudinal and quantum non-demolition (QND) coupling between the qubit and one quadrature of the cavity field. The measured qubit observable $\sigmaZ^\text{R}$ has an explicit time-dependence in the original interaction picture, $\sigmaZ^\text{R} \rightarrow \cos(\OmegaR t)\sigmaZ - \sin(\OmegaR t)\sigmaY$, making it analogous to a quadrature operator of a harmonic oscillator. The remaining terms in Eq.~\eqref{eqn:HRabi}, discussed explicitly in the supplemental material (SM) \cite{StrobeSupplement}, represent deviations from the ideal QND coupling, including terms that would cause an unwanted rotation of any input squeezing. These deleterious terms are rapidly oscillating, and hence are strongly suppressed if $\OmegaR \gg \kappa, \chi$. In contrast, there is no such suppression in a standard dispersive measurement, as such effects are resonant.

\begin{figure*}
  \includegraphics[width=\textwidth]{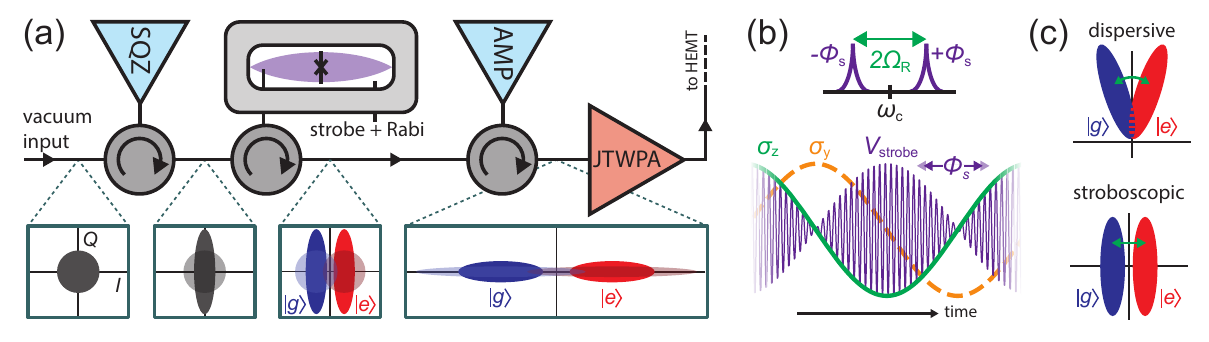}\\
  \caption{\label{intro_fig} (a) Simplified experimental setup. A first Josephson parametric amplifier (SQZ) injects squeezed vacuum into a superconducting cavity containing a qubit. The qubit is Rabi-driven at $\OmegaR$ about $\sigmaX$ and measured by two tones at frequencies $\omegaC \pm \OmegaR$, resulting in a qubit-state dependent displacement of the squeezed field in phase space. A second Josephson parametric amplifier (AMP) followed by a Josephson traveling wave parametric amplifier (JTWPA) perform phase-sensitive and phase-preserving amplification, respectively, of the output signal. Insets show phase-space representations of an ideal lossless measurement with (solid) and without (shaded) squeezing. (b) We choose the two tones' relative phase $\phi_s$ such that the envelope of the resulting measurement field is in phase with $\sigmaZ$. (c) Whereas dispersive readout rotates the output field in phase-space, stroboscopic readout displaces the output field, providing greater potential for enhancement by squeezing.}
\end{figure*}

To combine squeezing and stroboscopic measurement experimentally, we embed a 3D-transmon qubit \cite{Paik2011ObservationArchitecture} in a series configuration of two Josephson parametric amplifiers (JPAs) squeezing independent phase-space quadratures (\FigOne(a)). Several recent experiments not involving superconducting qubits have utilized similar configurations of superconducting amplifiers \cite{Mallet2011QuantumField, Flurin2012GeneratingLines,BienfaitMagneticMicrowaves}, which have been predicted to exhibit Heisenberg-like scaling in some regimes \cite{Yurke1986SU2Interferometers}. The qubit ($\omegaQ/2\pi = 3.898$ GHz) is coupled to a two-port superconducting waveguide cavity ($\omegaC/2\pi = 6.694$ GHz) with a dispersive interaction strength $\chi/2\pi = 0.73$ MHz. 

Into the weakly coupled port ($\kappa_\text{weak}/2\pi \leq \kappa_\text{int}/2\pi \sim 10$ kHz), we inject coherent qubit and cavity drives to generate stroboscopic measurement. The drive resonant with the qubit induces Rabi oscillations about $\sigmaX$ at $\OmegaR/2\pi = 40$ MHz exhibiting characteristic lifetimes $\TRabi \sim 20-30$ $\mu$s. Concurrently, a pair of cavity drives at frequencies $\omegaC \pm \OmegaR$, equivalent to a drive at $\omegaC$ modulated at $2\OmegaR$, stroboscopically probes the qubit state. Varying the relative phase of these two tones varies the timing of the modulation relative to the Rabi oscillations such that we can choose to measure any combination of $\sigmaY^\text{R}$ and $\sigmaZ^\text{R}$ (\FigOne(b)); for all measurements presented here we choose to measure $\sigmaZ^\text{R}$, equivalent to a $\sigmaZ$ measurement in the absence of a Rabi drive ($\OmegaR\rightarrow 0$). The measurement interaction (Eq.~\eqref{eqn:HRabi}) displaces the cavity output field in phase-space by $\pm \abar \chi / \kappa$ \cite{Hacohen-Gourgy2016QuantumObservables,StrobeSupplement}, in contrast to dispersive measurements which rotate the output field through the angle $\pm\arctan(2\chi/\kappa)$ or twice this angle for reflection measurements (\FigOne(c)).

Into the strongly coupled cavity port ($\kappa_\text{strong}/2\pi = 5.9$ MHz), we inject squeezed vacuum at $\omegaC$ produced by the first JPA, labeled SQZ in \FigOne(a). Keeping $\kappa_\text{strong} \gg \kappa_\text{weak}$ ensures that unsqueezed vacuum fluctuations incident to the weakly coupled port can be neglected and do not spoil the intracavity squeezing. We deliberately designed the squeezer to have a bandwidth smaller than $\OmegaR$ ($\kappa_{SQZ}/2\pi = 26$ MHz when the amplifier is off) to avoid generating significant squeezed noise power at the frequencies of the two measurement tones, similar to narrow-band squeezers used in previous works \cite{Toyli2016ResonanceVacuum}. We use a vector network analyzer to separately measure the phase-preserving squeezer gain, $\GSQZ$, from which we infer the amount of squeezing generated at this JPA. The resulting intracavity squeezed state is displaced in phase space from the origin along the $I$ quadrature by the stroboscopic measurement interaction (\FigOne(a)); we can freely adjust the squeezer pump phase to orient the squeezing angle $\Phi$ in phase-space parallel ($\Phi = 0$) or perpendicular ($\Phi = \pi/2$) to this signal.

In order for squeezing of vacuum fluctuations to have a significant effect on the SNR achieved at room temperature, the signal must be amplified with high efficiency. The signal travels from the cavity via the strongly-coupled port to a series of two superconducting amplifiers: a second JPA, labeled AMP in \FigOne(a), followed by a Josephson traveling wave parametric amplifier (JTWPA). The JTWPA functions as a high dynamic range amplifier ($P_{-1\text{dB}} \sim$ -100 dBm at 20 dB gain) that lowers the noise temperature of the measurement chain, referred to the JTWPA input, to be less than 1 K \cite{Macklin2015AAmplifier}. This permits operating the second JPA at modest gain less than 20 dB such that nonlinearities do not degrade the efficiency of the phase sensitive amplification \cite{BoutinEffectAmplifiers,BienfaitMagneticMicrowaves}.

%%%%%%%%%% SLOWING MEASUREMENT DEPHASING %%%%%%%%%%
This hardware configuration enables control via squeezing of the speed at which the cavity field acquires information about the qubit, which is reflected in the rate of measurement backaction. With the signal in the field quadrature $I$, backaction is exerted on the qubit by fluctuations of the conjugate variable $Q$. When no squeezing is applied, these fluctuations are those of the electromagnetic vacuum, which has a variance of 1/4 in all phase-space directions, and the resulting dephasing rate is given by $\GammaPhi = 2\bar{a}_0^2\chi^2/\kappa$ \cite{Hacohen-Gourgy2016QuantumObservables}. The observed decay of a Ramsey oscillation occurring during a measurement of a chosen strength indicates a steady-state dephasing rate $\GammaPhi = 0.54(1)$ $\mu\text{s}^{-1}$ (\FigTwo(a)). We compare to this baseline the $\GammaPhi$ seen when fluctuations in the measurement field are squeezed. Changing the amplitude of the JPA pump controls the amount of squeezing, while changing the pump phase rotates the squeezing relative to $I$ and $Q$. With the pump amplitude fixed such that $\GSQZ = 3.8$ dB, squeezing along the backaction quadrature $Q$ ($\Phi = \pi/2$) slows down $\GammaPhi$ by a factor of 1.8, indicating 2.5 dB of squeezing inside the cavity, and also amplifies fluctuations in the signal quadrature $I$, reducing the rate at which qubit state information leaves the cavity. The ability to slow a qubit measurement with squeezing is desirable in circumstances where a measurement occurs as an unwanted side-effect of a quantum operation, and has been proposed as a tool for realizing high-fidelity multi-qubit gates \cite{Puri2016High-FidelitySqueezing}. Conversely, squeezing along $I$ ($\Phi = 0$) increases $\GammaPhi$ by a factor of 3.9 (5.9 dB) and increases the output field SNR. We model the dephasing rate as
\begin{equation} \label{dephaseEqn}
\GammaPhi = \GammaPhiVac \frac{\Delta_Q^2}{1/4} = \GammaPhiVac (1+2\epsIn(N+M\cos2\Phi)),
\end{equation}
where $\Delta_Q^2$ is the field variance in the $Q$ quadrature inside the cavity, $1-\epsIn$ is the loss between the squeezer and qubit, and the squeezing parameters $N$ and $M$ \cite{Gardiner1986InhibitionSqueezing} are defined such that the variance of the amplified (squeezed) quadrature is $(1/2+N\pm M)/2$ at the squeezer output. We measure loss in the JPA and calculate a negligible effect on squeezing, so we model the JPA as producing an ideal squeezed state with $M=\sqrt{N(N+1)}$ and $N=\GSQZ-1$ both fixed by measurements of $\GSQZ$. A fit to the dephasing times shown in \FigTwo(b), jointly fit with the corresponding measurement times as discussed below, determines the input efficiency $\epsIn = 0.48$.
\begin{figure}
\includegraphics[width=\columnwidth]{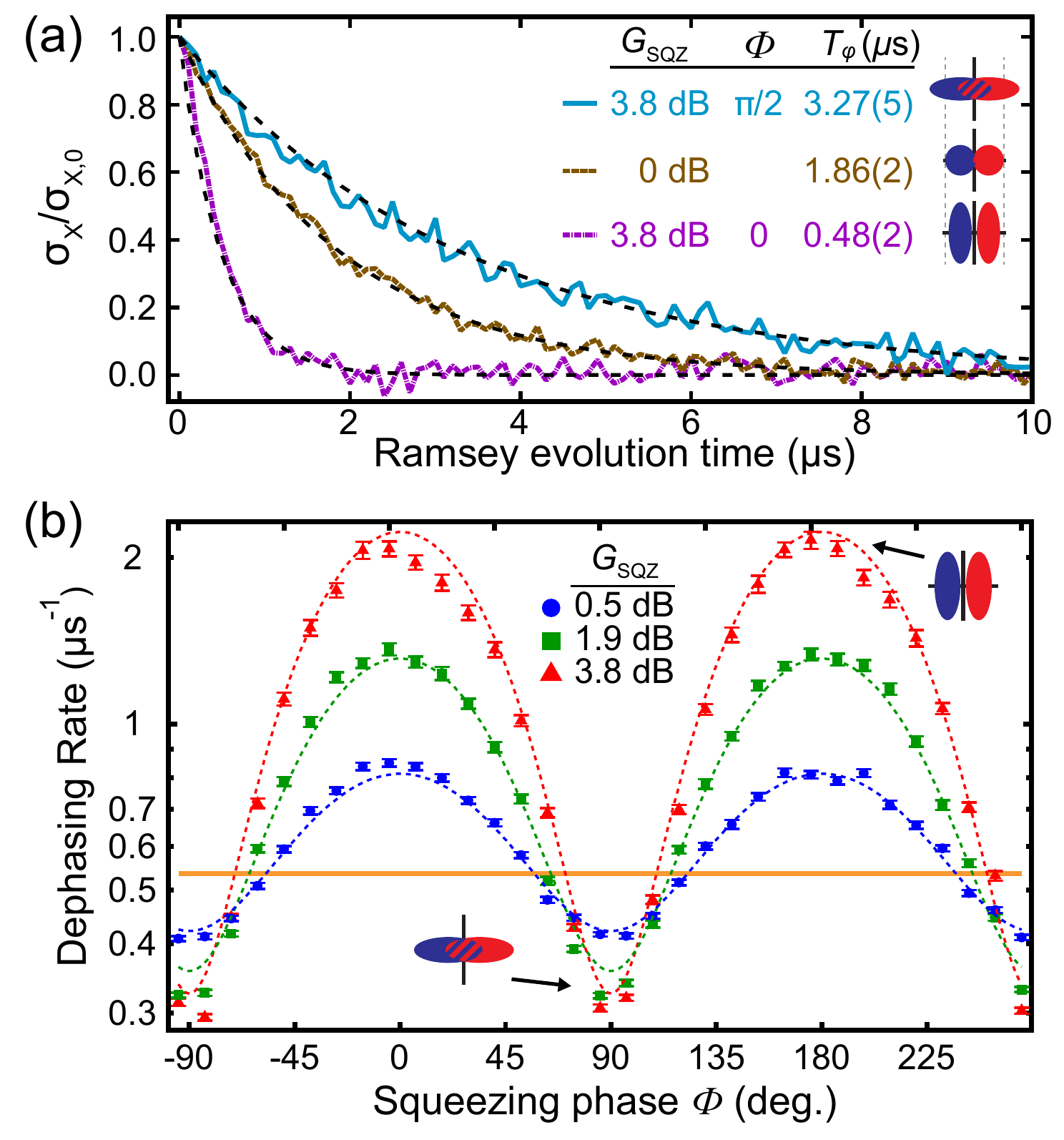}\\
\caption{\label{dephaseFig} (a) Ensemble-averaged Ramsey decay traces during a stroboscopic $\sigmaZ$ measurement with no squeezing (brown) or with squeezing in phase (purple) or out of phase (cyan) with the measurement signal. Each trace is normalized by its initial $t=0$ value, $\sigma_{x,0}$. Here the squeezer is pumped for 3.8 dB of phase-preserving gain as determined by Lorentzian fits of $S_{21}$ vs probe frequency. (b) Steady-state dephasing rates $\Gamma_\varphi = 1/T_\varphi$ were acquired via Ramsey measurements as in (a) repeated for a range of phases and squeezer gains. The orange horizontal band is the dephasing rate with the squeezer off. Error bars, including the width of the horizontal band, represent statistical fit uncertainties. The dashed curves are the results of a joint fit of all data in Figs. 2(b) and 3(b).}
\end{figure}

%%%%%%%%%% SPEEDING MEASUREMENT RATES %%%%%%%%%%
For a given squeezer setting, the squeezing-induced increase (decrease) in $\GammaPhi$ should correspond to an increase (decrease) in the SNR at our room-temperature homodyne detection setup. As in the dephasing rate study, we begin by determining a baseline with the squeezer off. We repeatedly prepare the qubit in either the ground or excited state and generate histograms of the results of stroboscopic measurements, indicated by circles in \FigThree(a). We calculate the SNR $= (2(\bar{V}_e-\bar{V}_g)/(\sigma_e+\sigma_g))^2$ from the mean separation and widths of the histogram distributions. For a given $\GammaPhiVac$ produced by an unsqueezed steady-state measurement field at the qubit, $\text{SNR} = 8 \GammaPhiVac \Tint \epsOut$, where $\Tint$ is the integration time and $\epsOut$ is the efficiency of the measurement chain downstream of the qubit. From the slope of SNR vs $\Tint$ we infer the steady-state measurement rate $\GammaMeasVac$ and the output efficiency $\epsOut = \GammaMeasVac / 2 \GammaPhiVac = 0.38$.

We repeat these measurements while squeezing or antisqueezing the noise in the signal quadrature, producing the histograms respectively indicated by stars and squares in \FigThree(a). With $\GSQZ = 4.0$ dB, we observe narrowed histograms with reduced overlap area; from the slope of SNR(t) we determine a 24\% increase in $\GammaMeas$ from 0.41(1) without squeezing to 0.51(1) $\mu s^{-1}$ with squeezing. Additional measurements display the dependence of $\GammaMeas$ on squeezing amount and phase, as shown in \FigThree(b). As the SNR and thus $\GammaMeas$ should vary inversely with variance $\Delta_I^2$ at the end of the measurement chain, we fit the data with the expression
\begin{align}
\begin{split} \label{measEqn}
\GammaMeas &= \GammaMeasVac \frac{1/4}{\Delta_I^2} \\
=& \GammaMeasVac (1+2\epsIn\epsOut(N-M\cos2\tilde{\Phi}))^{-1}.
\end{split}
\end{align}
The free parameters in the joint fit of $\GammaPhi$ and $\GammaMeas$ are $\epsIn = 0.48$, a global phase, and an offset $\delta = \tilde{\Phi} - \Phi = 14\degr$ capturing imperfect alignment of AMP with the signal quadrature, which shifts $\GammaMeas(\Phi)$ but not $\GammaPhi(\Phi)$. We fix $\GammaPhiVac$, $\GammaMeasVac$, and $\epsOut$ at the values found with no squeezing. As expected, $\GammaMeas(\Phi)$ is $\pi$-periodic, with phases maximizing $\GammaMeas$ close to those maximizing $\GammaPhi$.
\begin{figure}
\includegraphics[width=\columnwidth]{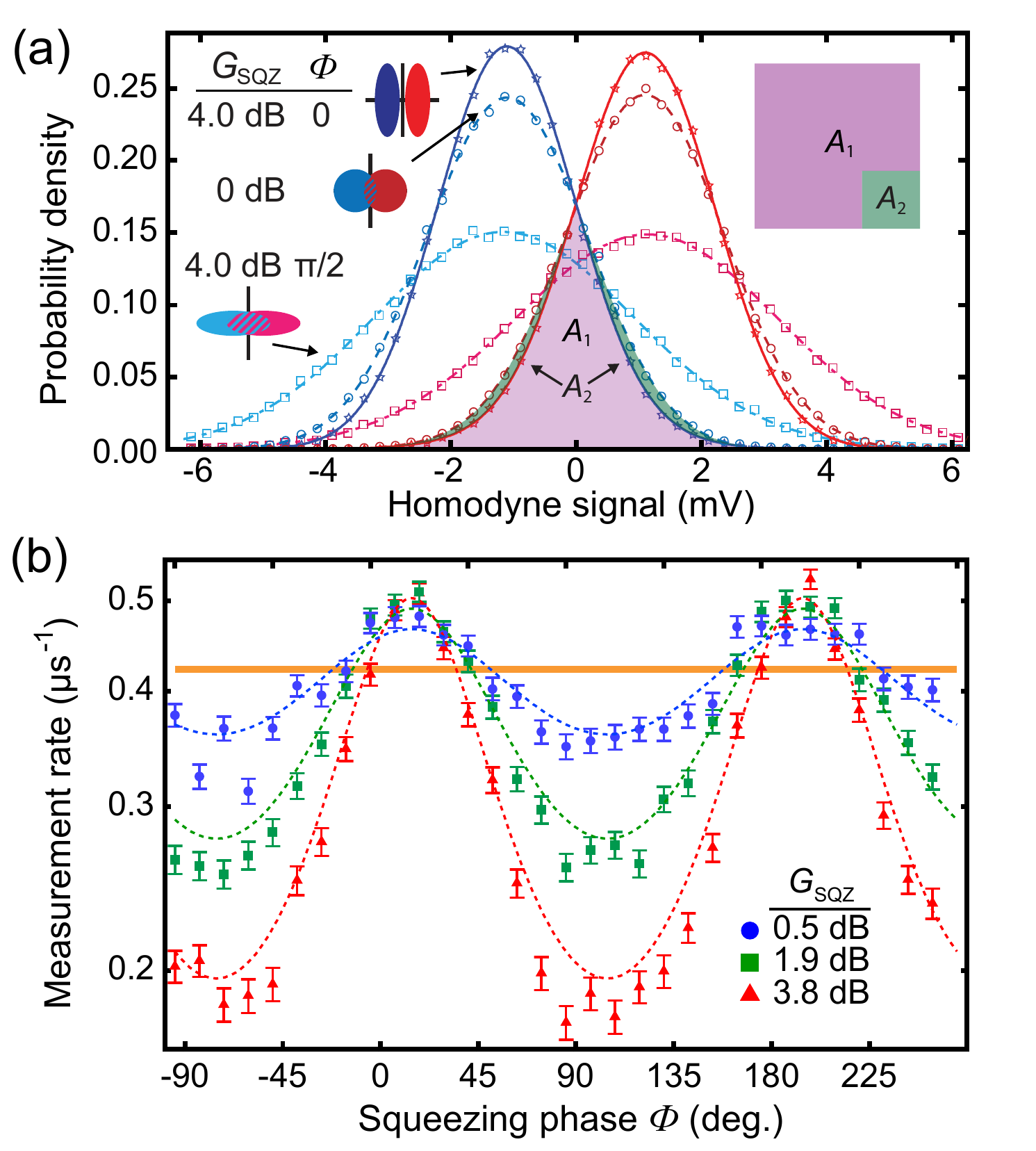}\\
\caption{\label{measFig}(a) Datapoints are normalized histograms of the mean homodyne voltage integrated for 1.8 $\mu s$ conditioned on preparing the qubit in the ground (blue colors) or excited (red colors) states. Curves are Gaussian fits. Measurements were repeated with squeezed (stars, solid), unsqueezed (circles, dashed) and antisqueezed (squares, dot-dashed) noise. Fits of data conditioned on excited-state preparation include a small second Gaussian to capture ground-state population ($\lesssim 2\%$) attributable to qubit relaxation before and during the Rabi ramp-up. The overlap area is smaller with squeezing ($A_1$) than without squeezing ($A_1+A_2$). (b) The measurement rate $\GammaMeas$ was determined for multiple phases and amounts of squeezing. The orange horizontal band is $\GammaMeas$ with SQZ off. Error bars, including the width of the horizontal band, are standard errors of fits of SNR(t). The dashed curves are results of a joint fit of all the data in Figs. 2(b) and 3(b) with free parameters $\epsIn$, $\delta$, and a global phase.}
\end{figure}

%%%%%%%%%% ENHANCING EFFICIENCY %%%%%%%%%%
Comparing $\GammaPhi$ and $\GammaMeas$ reveals that, despite the intrinsic fragility of squeezing in the presence of loss, it is also possible for injected squeezing to improve the measurement efficiency $\eta$ in a lossy environment. Here we define $\eta = \GammaMeas/2\GammaPhi$ such that for perfect efficiency $\eta = 1$. With the squeezer off, the efficiency reduces to $\eta_\text{vac} = \epsOut$, set by loss and added noise in the measurement chain downstream of the qubit. With the squeezer on, $\eta$ depends also on $\epsIn$, according to
\begin{equation}\label{etaVsPhase}
\eta = \frac{\GammaMeas}{2\GammaPhi} = \eta_{\text{vac}}\frac{\GammaPhiVac}{\GammaPhi}\frac{\GammaMeas}{\GammaMeasVac}, 
\end{equation}
where the ratios are manifest in equations \ref{dephaseEqn} and \ref{measEqn}, respectively. \FigFour shows $\eta(\Phi)$, calculated by dividing Fig. 3(b) by Fig. 2(b); near $\Phi=\pi/2$, $\eta$ increases from 0.38(1) to 0.42(1). The increase can be understood by comparing the effect of the output loss $1-\epsOut$ on SNR with and without squeezing. In both cases, the mean signal size at the cavity output is the same, and this signal is attenuated by the factor $1-\epsOut$. Without squeezing, the vacuum fluctuations in the signal quadrature $I$ are unaffected by the loss, so the SNR is reduced by $1-\epsOut$. In contrast, squeezing along $Q$ ($\Phi = \pi/2$) injects amplified noise in $I$ which does get attenuated, partially canceling the effect of loss on SNR. Although the initial SNR leaving the cavity is lower in this case, so is the measurement backaction, as the $Q$ quadrature is squeezed. Thus at the cost of decreasing $\GammaMeas$, $\eta$ can be increased, with greater enhancement seen in systems with $\epsIn \gg \epsOut$. For example, $\eta>50\%$ with only phase-preserving amplification of the signal is possible. Recently, a similar technique was demonstrated using squeezing to increase robustness of optical cat-states \cite{LeJeannicSlowingSpace}. The orthogonal case ($\Phi=0$) has the reverse effect, increasing $\GammaMeas$ and decreasing $\eta$, with more speed-up in systems with high $\epsOut$. 
\begin{figure}
  \includegraphics[width=\columnwidth]{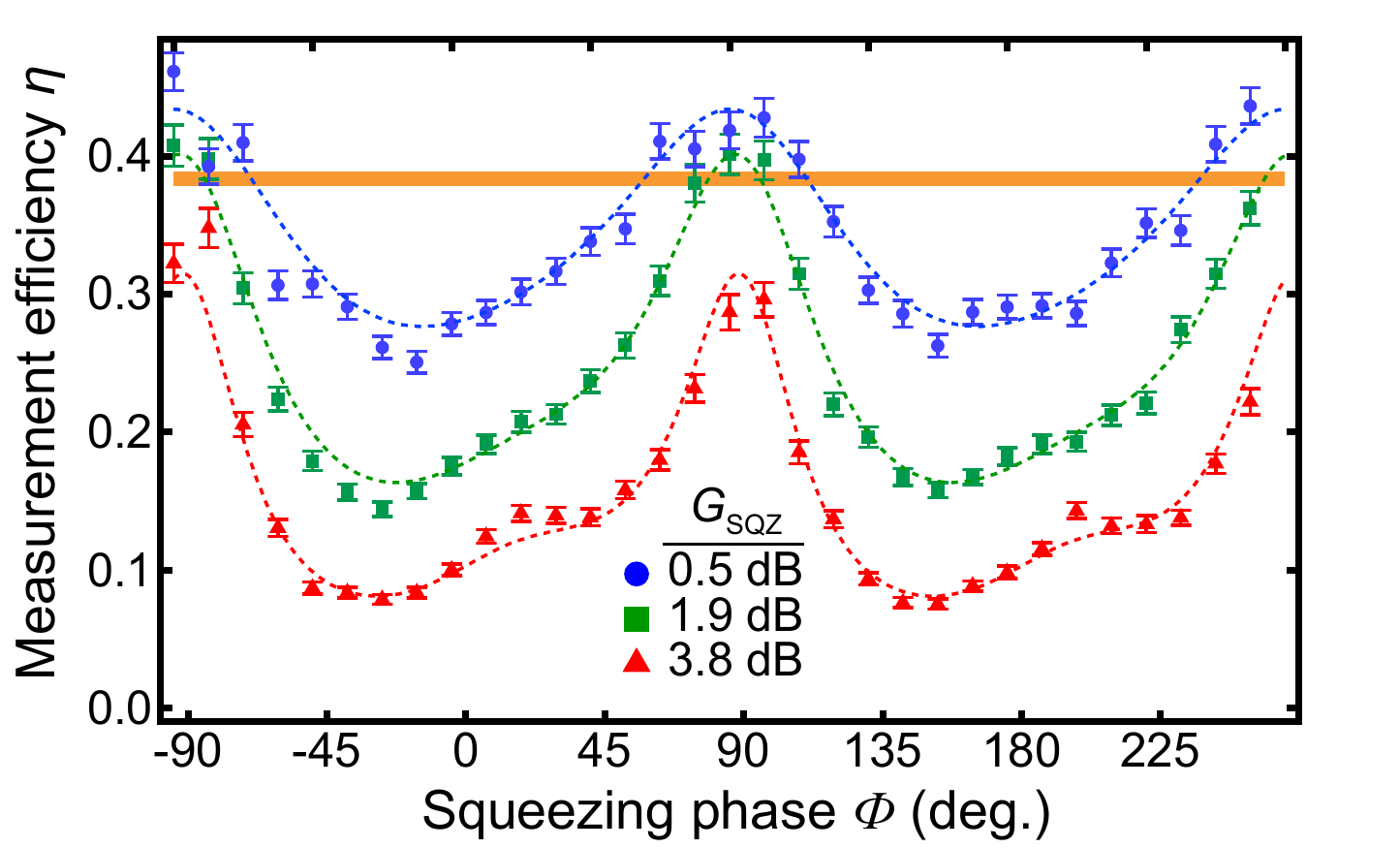}\\
  \caption{\label{etaFig} The overall measurement efficiency $\eta(\Phi)$ is shown for several values of $\GSQZ$. The data and fits are computed from the corresponding elements in Figs. 2(b) and 3(b). We attribute the asymmetry about phases where $\Phi$ is a multiple of $\pi/2$ to misalignment of the signal quadrature with the amplification quadrature of the second JPA. A small increase in $\eta$ above $\epsOut$ can be resolved at low squeezer gain near $\Phi = \pi/2$.}
\end{figure}

%%%%%%%%%% CONCLUSION %%%%%%%%%%
In summary, this work demonstrates that input squeezing can reduce the noise in measurements of standard superconducting qubits, and can also slow a measurement process and its associated dephasing. In general, the photon number is bounded by nonlinearity of the qubit-field interaction, so maximizing information per photon is useful for \textit{e.g.} speeding up a quantum algorithm in which measurement is a bottleneck. Similarly, by squeezing the other quadrature, it may be possible to reduce the necessary wait time for cavity depopulation at the end of a measurement. A natural next step is to implement our techniques in a system highly optimized for efficiency \cite{Walter2017RapidQubits}, possibly incorporating ongoing development of superconducting circulators or on-chip amplifiers \cite{Sliwa2015ReconfigurableAmplifier,Kerckhoff2015On-ChipRotation,LevitanThesis,Metelmann2015NonreciprocalEngineering}. As squeezing cannot improve the SNR by more than a factor of $(1-\epsOut)^{-1}$, here equal to 1.6, efficiency improvements would better leverage the large amounts of microwave squeezing, exceeding 12 dB \cite{Eichler2014Quantum-LimitedResonators}, produced to date. Absent loss, our experimental setup is predicted to effectively utilize up to 16 dB of squeezing \cite{StrobeSupplement}. Complementary studies may explore the limits of stroboscopic readout speed by optimizing couplings or by increasing $\OmegaR$ to further suppress counter-rotating terms. The latter may be facilitated by modern high-anharmonicity qubit designs such as the C-shunt flux qubit \cite{Yan2016TheReproducibility}. Finally, it would be useful to investigate transient effects in the presence of squeezing, which are of increasing importance for shorter measurement times. Here stroboscopic or other longitudinal readout schemes may be advantageous even without squeezing, as the cavity-field ring-up and ring-down trajectories are expected to follow straight lines in phase space in contrast to the circuitous short-time response of dispersive measurements \cite{Didier2015FastInteraction}.

%%%%%%%%%% ACKNOWLEDGEMENTS %%%%%%%%%%
The authors thank E. Flurin for useful discussions, and MIT Lincoln Labs for fabrication of the JTWPA. Work was supported by the Army Research Office (W911NF-14-1-0078). A.E. acknowledges support from the Department of Defense through the NDSEG fellowship program. L. M. acknowledges support from the Berkeley Fellowship and the National Science Foundation Graduate Research Fellowship.

%%%%%%%%%% Merge with supplemental materials %%%%%%%%%%
\pagebreak
\widetext
%%%%%%%%%% Prefix a "S" to all equations, figures, tables and reset the counter %%%%%%%%%%
\setcounter{equation}{0}
\setcounter{figure}{0}
\setcounter{table}{0}

\clearpage

\newcommand{\FigSOne}{Fig.~\ref{detailedSetup}}
\newcommand{\FigSTwo}{Fig.~\ref{snrPlots}}
\newcommand{\FigSThree}{Fig.~\ref{fig:theoryPlots}}
\renewcommand{\thefigure}{S\arabic{figure}}
\renewcommand{\Re}{\operatorname{Re}}
\renewcommand{\Im}{\operatorname{Im}}

\begin{center}
{\large \bf Stroboscopic qubit measurement with squeezed illumination: Supplemental material}
\end{center}

\section{Details of the experimental setup}
A detailed diagram of our experimental setup is shown in \FigSOne. The circuit is driven by three microwave generators. The output of a generator near the qubit frequency is modulated to produce qubit pulses. A second generator provides the pump for the Josephson traveling wave parametric amplifier (JTWPA). The output of the third generator, at the cavity frequency, is split to power five distinct phase-locked operations, detailed below.

Two of the five splitter outputs are frequency doubled; these are used to flux-pump the two Josephson parametric amplifiers (JPAs) in the circuit. The phase and amplitude of the squeezer pump is controlled by a digital phase-shifter and voltage variable attenuator, allowing us to automatically sweep the squeezing phase while keeping the pump power (sampled by a spectrum analyzer) constant. Low-pass filters prevent pump power leakage through the JPAs from driving the qubit or JTWPA. To prevent pump power from the second JPA from dephasing the qubit via higher cavity modes, pump power is tapped off at room temperature and sent to the cavity as a cancellation tone. The JPAs themselves are flux-biased downward from their maximum frequencies by off-chip superconducting coils (not shown), allowing for flux pumping. 

A third splitter output acts as the local oscillator (LO) for an IQ mixer used to generate stroboscopic readout pulses. The I and Q ports are driven simultaneously by 40 MHz pulses sourced by an arbitrary waveform generator (AWG). The phase of the 40 MHz pulse on the I port is varied by a voltage-variable IF phase-shifter; we use this phase to balance the amplitudes of the two sidebands output by the mixer.

The fourth splitter output drives another IQ mixer to produce dispersive readout pulses. These are used for diagnostic purposes and ground-state heralding.

The final splitter output provides the LO for the IQ mixer used for homodyne demodulation of the signal leaving the fridge. The demodulated signal is further amplified and digitized via standard techniques.
\begin{figure}[!h]
\includegraphics[width=0.8\textwidth]{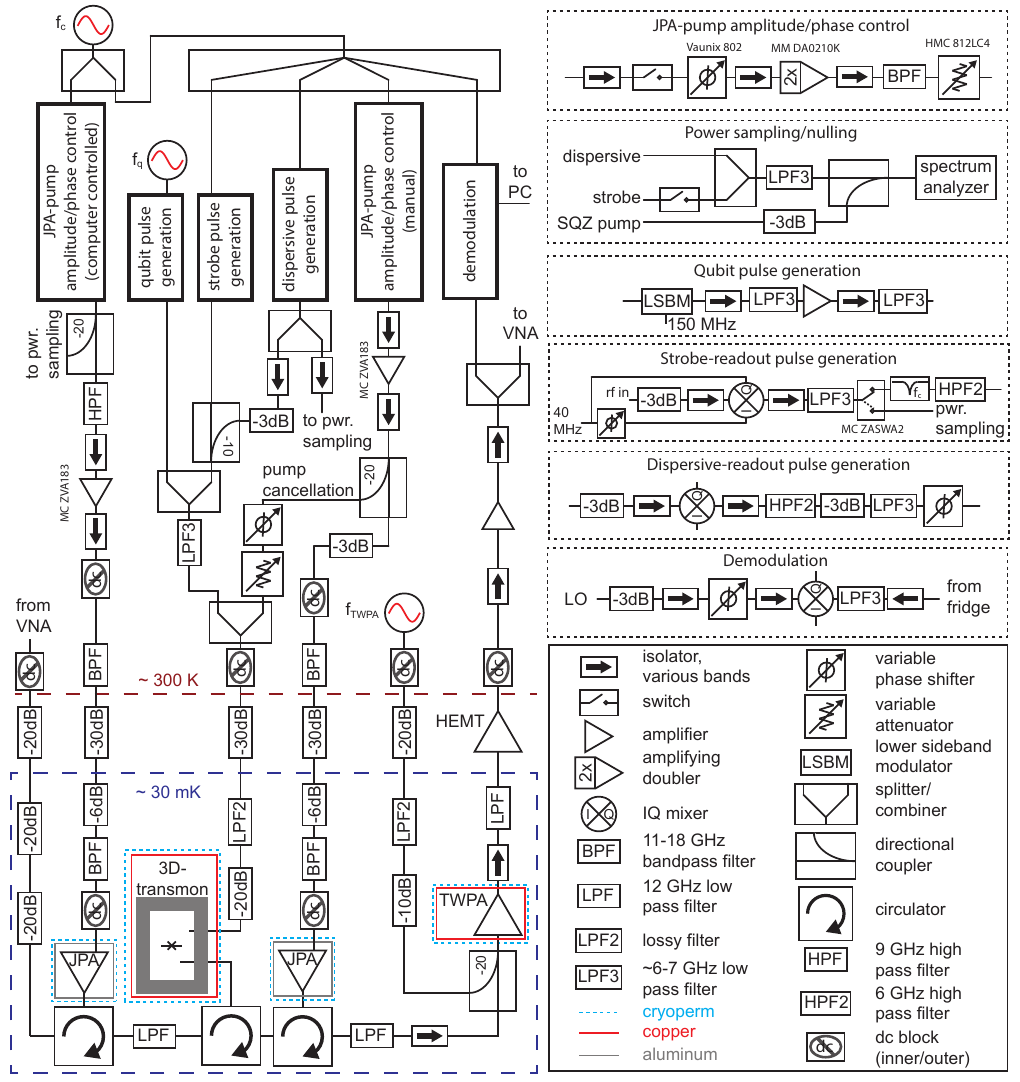}\\
\caption{\label{detailedSetup} Detailed experimental setup}
\end{figure}

\section{Stroboscopic Readout Calibration}
The calibration procedure for setting up the stroboscopic readout is extensive but automated. Much of the procedure follows that given in the Methods section of \cite{Hacohen-Gourgy2016QuantumObservables}, which should be consulted for additional details. Switches and frequency multiplexing allow for automatic nulling of both readout mixers and leveling of the squeezer power using one spectrum analyzer. The strobe mixer is nulled with the IF modulation applied. The Stark shift in the presence of the strobe tones and injected squeezed noise is determined via Ramsey measurements using dispersive readout. A range of Rabi drive powers and a range of phases of the 40 MHz modulation sourced by the AWG are both swept to align the strobe timing with the 40-MHz oscillations of $\sigmaZ$ (Fig. 1(b)), and pulse sequences are automatically generated as needed. All sequences utilizing stroboscopic readout contain intercalary measurements of Rabi oscillations, the results of which are used to stabilize the Rabi frequency to 40 MHz on timescales shorter than ambient temperature drifts. We perform tomographic measurements of Ramsey decays to balance the sideband amplitudes at the qubit. In the case of balanced sideband amplitudes, the transverse component of the Bloch vector, which we call $\sigmaX$, will decay to zero; imbalanced sidebands produce cooling effects which result in a non-zero steady-state $\langle\sigmaX\rangle$. We adjust the phase-shift on the I port of the strobe mixer to minimize this quantity, balancing the sidebands.

\section{Superconducting Devices}
The qubit is a 3D-transmon, fabricated by a standard aluminum lift-off process on silicon,  contained in an aluminum cavity resonator. The two JPAs are both lumped-element resonators fabricated via aluminum-liftoff on silicon. The designs are similar to that shown in \cite{Toyli2016ResonanceVacuum}. Both devices include multiple SQUIDs and geometric meander inductance to weaken the device nonlinearity, which is expected to reduce nonidealities in the squeezing performance. The squeezer is made narrowband compared to the strobe-tone separation (40 MHz) to avoid introducing significant squeezed noise power at the strobe-tone frequencies. The strobe separation frequency cannot be made arbitrarily large, as larger detunings make it more difficult to couple measurement power into the cavity. The second JPA is more broadband, which increases its dynamic range but also increases the pump power required. In the present setup, heating due to this pump places a soft limit on the phase-sensitive gain and thus, likely, also on $\epsOut$. We expect simply removing attenuation from the pump line for the second JPA would enable slightly greater $\epsOut$ and effect of squeezing on $\GammaMeas$. The JTWPA was fabricated by MIT-Lincoln Laboratories using a niobium trilayer process on silicon. This highly broadband device was found to be sensitive to leakage of the high-frequency flux-pump of the JPA; a low-pass filter (K\&L Microwave L-series) eliminated this problem. The device remained highly stable over the duration of the project. Performance details of similar devices are given in \cite{Macklin2015AAmplifier,Toyli2016ResonanceVacuum}.

For each dataset, a vector network analyzer (VNA) was used to determine the squeezer gain. The measurement bandwidth was kept small compared to the detuning of the VNA signal, such that the inferred power gains $\GSQZ$ correspond to phase-preserving amplification. Assuming ideal squeezing, one can find the phase-sensitive power gain and the amount of squeezing by the relation $1/2+N\pm M = \frac{1}{2}\left(\sqrt{\GSQZ} \pm \sqrt{\GSQZ-1} \right)^2$. VNA characterization of $\GSQZ$ was performed before and after sets of measurements to confirm squeezer stability.
% Comment: Accounting for the small loss in SQZ (kappa_int ~ 122 kHz) changes inferred N, M, and N-M values by less than 1 percent).

\section{Bounding the Thermal Photon Number}
The term squeezing implies not only reduction of a quadrature noise variance, but specifically reduction below the vacuum variance. Thus an observed reduction in noise can only be called squeezing once the initial electromagnetic field is confirmed to have a sufficiently small thermal component. Since thermal photons dephase the qubit, one can bound the mean thermal photon number using the qubit dephasing time and the dispersive qubit-cavity coupling $\chi/2\pi = 0.73$ MHz. Using a 100 $\mu$s echo sequence with a single $\pi$-pulse, we found $T_2 = 64(3)$ $\mu$s. Because the evolution times before and after the $\pi$ pulse are very long compared to the cavity response time $1/\kappa$, the effect of the echo on the contribution of thermal cavity photons to the qubit dephasing is negligible. Solving 
\begin{equation}
(T_2)^{-1} = \frac{\kappa}{2}\Re\left[\sqrt{\left(1+\frac{2i\chi}{\kappa}\right)^2+\left(\frac{8i\chi n_\text{th}}{\kappa}\right)}-1\right]
\end{equation}
bounds the effective thermal photon number $n_\text{th} \leq .01$, where equality would imply the dephasing is exclusively due to the presence of thermal photons \cite{Clerk2007UsingOscillator}.

\section{Determining the Measurement Rate}
Measurement rates were determined by repeatedly preparing the qubit in either the ground or excited state and applying the stroboscopic measurement tones simultaneous with a Rabi drive. Once transient ring-up behaviors had become small (steady-state), the output homodyne voltage was digitized with a sampling rate of 20 MS/s for 1.8 $\mu$s. A dispersive readout pulse was used to herald the ground state prior to state preparation and measurement. Measurement records corresponding to times during which the Rabi frequency was estimated to differ from 40 MHz by more than 10 kHz were rejected. No further post-selection was performed on the data. The voltage records were used to calculate the ensemble-mean value and the ensemble-standard deviation of the integrated voltage signal as a function of time. As expected, the separation of the ensemble-mean voltages conditioned on the ground and excited states remained approximately constant with integration time, while the standard deviation of each ensemble decreased. Plotting the square of the ratio of these quantities shows an approximately linear dependence of the SNR on integration time. A linear fit gives the measurement rate; we choose a sampling rate slow compared to the decay of the ensemble-averaged autocorrelation function to minimize correlations in the statistical fluctuations of subsequent samples, which might otherwise artificially reduce the standard error of the linear fit.
\begin{figure}[!h]
\includegraphics[width=0.7\textwidth]{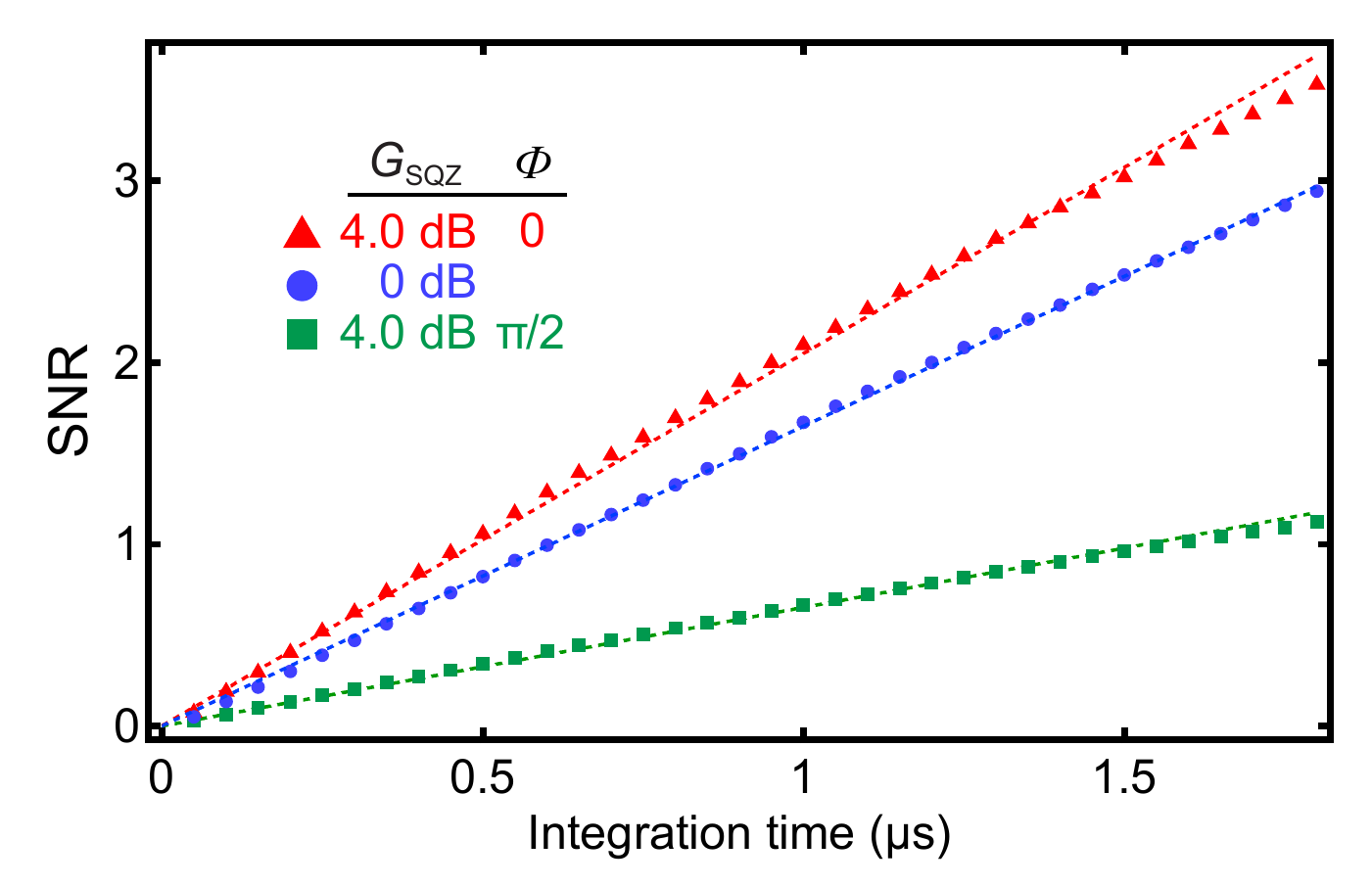}\\
\caption{\label{snrPlots} Representative plots of SNR vs integration time}
\end{figure}

\FigSTwo \ shows representative plots of power SNR vs time for squeezed, unsqueezed, and antisqueezed stroboscopic measurements. The same data were used to generate Fig. 3(a). The linear fits capture the time dependence reasonably well, though at longer times the SNR grows sublinearly; this behavior is not fully understood at present, perhaps relating to counter-rotating terms in the measurement Hamiltonian.

%%%%%%%%%%%%%%%%%%%%%%%%%%%%%%% THY 1 %%%%%%%%%%%%%%%%%%%%%%

\renewcommand{\labelitemi}{$\bullet$}

\section{Effective longitudinal Hamiltonian}

In this section, we briefly show how to obtain the effective longitudinal Hamiltonian introduced in
Eq. (1) of the main text.

The Hamiltonian describing the system is given by
\begin{equation}
	\hH (t) = \hH_{\mm{JC}} + \hH_{\mm{Q}} (t) + \hH_{\mm{drive}} (t),
	\label{eq:Hmeasurement}
\end{equation}
where $\hH_{\mm{JC}}$ is the Jaynes-Cummings Hamiltonian
\begin{equation}
	\hH_{\mm{JC}} = \frac{1}{2} \omega_\uq \sz + \omega_\uc \ha^\dag \ha + g\left(\ha \sp +
	\ha^\dag \sm\right),
	\label{eq:HJaynesCummings}
\end{equation}
with $\omega_\uq$ the qubit angular frequency, $\omega_\uc$ the cavity angular frequency, and $g$ the qubit-cavity coupling strength. The Pauli operators are given by $\hat{\sigma}_j$, $j \in
\{x,y,z\}$, and the photon annihilation (creation) operators by $\ha$ $\left(\ha^\dag\right)$.

During the measurement, the qubit is driven by
\begin{equation}
	\hH_{\mm Q} (t) = \Omega_\uR \cos\left[ \omega_\ud t \right] \sx,
	\label{eq:HRabi}
\end{equation}
which causes the qubit to undergo Rabi oscillations. Here, $\Omega_\uR$ denotes the amplitude of the
drive while $\omega_\ud$ denotes the angular frequency of the drive.

In addition to the qubit drive, there is also a two-tone drive applied to the cavity with
Hamiltonian:
\begin{equation}
	\begin{aligned}
		\hH_{\mm{drive}} (t) &= \left(\varepsilon_+ \exp[-i (\omega_\uc + \Omega_\uR) t] +
		\varepsilon_- \exp[-i (\omega_\uc - \Omega_\uR) t]\right) \ha^\dag + h.c.
	\end{aligned}
	\label{eq:HTwoTone}
\end{equation}
The amplitude of the drives are given by $\varepsilon_\pm$ and they are detuned by $\pm \Omega_\uR$
relative to the cavity angular frequency.

Given that the experiment is performed in the dispersive regime, i.e. $\Delta=\omega_\uq -
\omega_\uc \gg g$, we can (as is standard) move to the dispersive frame (i.e. make a Schrieffer-Wolff transformation to eliminate the Jaynes-Cummings interaction to leading order).  After this transformation (and working in an interaction picture with respect to free cavity and qubit Hamiltonians) we have
\begin{equation}
	\begin{aligned}
		\hH_{\mm{eff}} (t) &= \frac{g^2}{\Delta}\left( \ha^\dag \ha +
		\frac{1}{2}\right) \sz \\
		&\phantom{={}} + \frac{1}{2}\Omega_\uR \left(\exp[i \omega_\ud t] + \exp[-i
		\omega_\ud t] \right) \left( \exp[i \omega_\uq t] \sp + \exp[-i \omega_\uq t]
		\sm\right)\\
		&\phantom{={}} + \left(\varepsilon_+ \exp[-i \Omega_\uR t] + \varepsilon_- \exp[i
		\Omega_\uR t]\right) \ha^\dag \\
		&\phantom{={}} + \left(\varepsilon_+^\ast \exp[i \Omega_\uR t] +
		\varepsilon_-^\ast \exp[- i \Omega_\uR t]\right) \ha \\
		&\phantom{={}} + \mathcal{O}\left(\frac{g^3}{\Delta^2}, \frac{g}{\Delta}\Omega_\uR,
		\frac{g}{\Delta} \varepsilon_\pm\right),
	\end{aligned}
	\label{eq:HeffSW1}
\end{equation}
where we have omitted terms proportional to the identity.

Higher order terms in Eq. \eqref{eq:HeffSW1} can be safely ignored for a large enough detuning $\abs{\Delta}$.  Assessing the impact of these terms perturbatively, one finds that the dispersive approximation is a good description as long as the following conditions hold:
\begin{equation}
	\begin{aligned}
		\frac{g\Omega_\uR}{\Delta^2} &\ll 1,\\
		\frac{g\Omega_\uR}{\Delta(\Delta + 2\omega_\uc)} &\ll 1,\\
		\frac{g}{\Delta}\frac{\abs{\varepsilon_+}}{\abs{\Omega_\uR - \Delta}} &\ll 1,\\
		\frac{g}{\Delta}\frac{\abs{\varepsilon_-}}{\abs{\Omega_\uR + \Delta}}
		&\ll 1.
	\end{aligned}
	\label{eq:condValSW}
\end{equation}
Given that experimentally we have $\omega_\uc/2\pi = 6.694\,\mm{GHz},\ \omega_\uq/2\pi = 3.898\,\mm{GHz},
\Omega_\uR/2\pi = 40\,\mm{MHz},\ \abs{\Delta} = 2.796\,\mm{GHz}$, and $g = 45.2\,\mm{MHz}$, the above conditions are well-fulfilled.

We next proceed with the displacement transformation introduced in the main text: $\ha \to 2
\bar{a}_0\cos\left(\Omega_\uR t\right) + \hd$. This yields (in the rotating frame at $\omega_{\rm c}$ for the cavity and $\omega_{\rm d}$ for the qubit)
\begin{equation}
	\begin{aligned}
		\hH_{\mm{eff}} (t) &= \frac{1}{2}\Delta_{\rm q}(t)\sz + \chi \hd^\dag \hd \sz + \chi \bar{a}_0 \cos\left(\Omega_\uR t\right)
		\left( \hd + \hd^\dag \right) \sz + \frac{1}{2}\Omega_\uR \sx,
	\end{aligned}
	\label{eq:HeffFinal}
\end{equation}
where we have introduced $\chi = g^2/\Delta$ and performed the RWA approximation, $\Omega_\uR/ 2
\tilde{\omega}_\uq \ll 1$ and $\abs{\varepsilon_\pm}/\Omega_\uR \ll 1$. The time dependent qubit detuning
\begin{align}
	\Delta_{\rm q}(t) = \omega_{\rm q} + \chi + 4\chi\bar{a}_0^2 \left[1 + \cos\left(2 \Omega_\uR t\right)\right] - \omega_{\rm d},
\end{align}
contains static shifts coming from both the dispersive interaction and the average cavity amplitude, as well as a time dependent shift due to the oscillatory nature of the cavity displacement. The static shifts can be accounted for by appropriately choosing $\omega_{\rm d} = \omega_{\rm q} + \chi + 4\chi\bar{a}_0^2$. Using a standard unitary transformation, the time dependent shift can be gauged away, resulting in a renormalization of the Rabi amplitude on the order of $(\chi \bar{a}_0^2/\Omega_{\rm R})^2$, as well as time dependent sideband qubit drives. From here on, we assume that $\Omega_{\rm R}$ represents the renormalized Rabi amplitude, though for the parameters of the experiment this renormalization is too small to be relevant. We can safely ignore the sideband terms as the amplitude of the oscillations they generate is smaller than the line-width of the qubit, such that they will have no noticeable effect on the qubit dynamics.

In the frame rotating at the Rabi frequency $\Omega_\uR$ for the qubit (Rabi frame), the Hamiltonian is
\begin{equation}
	\begin{aligned}
		\hH_\uR (t) &= \chi \bar{a}_0\hat{\sigma}^{\rm R}_z\left(\hat{d} + \hat{d}^\dagger\right)
	  	+ \left(e^{i\Omega_Rt}\hat{A} + e^{i2\Omega_Rt}\hat{B} + h.c. \right),
	\end{aligned}
	\label{eq:HBAE}
\end{equation}
where
\begin{align}
	&\hat{A} = \frac{\chi}{2}\hd^\dag \hd \left(\sz^{\rm R} - i\sy^{\rm R}\right), \\
	&\hat{B} = \frac{\chi\bar{a}_0}{2}\left(\hd + \hd^\dag\right)\left( \sz^{\rm R} - i \sy^{\rm R}\right).
\end{align}
The first term of Eq.~\eqref{eq:HBAE} is the ideal synthetic longitudinal Hamiltonian that allows
for a quantum non-demolition (QND) measurement. The remaining terms lead to imperfections of the measurement
scheme. The term proportional $\hd^\dag \hd \sz^{\rm R}$ leads to a rotation of the input squeezing and
limits the maximum useful amount of input squeezing. The terms proportional to $\sy^{\rm R}$ induce
spin-flips and therefore constitute non-QND imperfections.

%%%%%%%%%%%%%%%% THEORY SECTION 2 $%%%%%%%%%%%%%%%%%%%%
\section{Signal-to-Noise Ratio}

In this section, we show that, in contrast to the standard dispersive setup, unwanted terms in our synthetic longitudinal scheme (i.e. $\hat{A}$ and $\hat{B}$ in Eq.~\eqref{eq:HBAE}) only lead to a small undesirable qubit-dependent rotation of an inputted squeezed state. As a result, with the current experiment, one could advantageously use $>10$ dB of squeezing to enhance a measurement without being limited by unwanted effects. To show this, we calculate the signal-to-noise ratio (SNR), assuming the cavity is driven with squeezed vacuum, and that one quadrature of the output field leaving the cavity is measured via a homodyne setup.

We start by neglecting all non-QND imperfections, i.e. we treat the qubit as a classical variable
($\avg{\sz^{\rm R}} = \bar{\sigma}_z = \pm 1,\, \avg{\sy^{\rm R}} = 0$). Within this framework, the Langevin equation for $\hd (t)$ is
\begin{equation}
		\rd_t \hd (t) = -i \chi \bar{a}_0 \left(1 + \cos\left[2 \Omega_\uR t\right]\right) \bar{\sigma}_z - i \chi
		\cos\left(\Omega_\uR t\right) \bar{\sigma}_z \hd  - \frac{\kappa}{2} \hd (t) -
		\sqrt{\kappa} \hd_{\mm{in}} (t).
	\label{eq:Langevind}
\end{equation}

Equation~\eqref{eq:Langevind} can be straightforwardly integrated and yields
\begin{equation}
	\begin{aligned}
		\hd (t) &= -2 i \beta_z \Omega_\uR \bar{a}_0 e^{-\frac{\kappa}{2} t} \exp\left[-i
		\beta_z \sin\left(\Omega_\uR t\right)\right] \int_{-\infty}^t \ud t_1
		\exp\left[\frac{\kappa}{2} t_1\right] \exp\left[i \beta_z \sin\left(\Omega_\uR
		t_1\right)\right] \cos^2\left(\Omega_\uR t_1\right) \\
		&\phantom{={}}
		+\sqrt{\kappa} e^{-\frac{\kappa}{2} t}
		\exp\left[-i \beta_z \sin\left(\Omega_\uR t\right)\right]\int_{-\infty}^t \ud t_1
		\exp\left[\frac{\kappa}{2} t_1\right] \exp\left[i \beta_z
		\sin\left(\Omega_\uR t_1\right)\right]\hd_{\mm{in}} (t_1),
	\end{aligned}
	\label{eq:d}
\end{equation}
where we have defined
\begin{align}
	\beta_z = \frac{\chi}{\Omega_\uR} \bar{\sigma}_z \label{eq:Bz}
\end{align}
and $\hd_{\mm{in}} (t)$ is the
(squeezed) input field defined by its statistics
\begin{equation}
	\begin{aligned}
		\avg{\hd_{\mm{in}} (t)} &= 0,\\
		\avg{\hd_{\mm{in}}^\dag (t) \hd_{\mm{in}} (t')} &= \sinh^2(r) \delta(t-t'), \\
		\avg{\hd_{\mm{in}} (t) \hd_{\mm{in}} (t')} &= \frac{1}{2} \sinh (2r)
		e^{2 i \Phi} \delta(t-t')\\
		\avg{\hd_{\mm{in}}^\dag (t) \hd_{\mm{in}}^\dag (t')} &= \frac{1}{2} \sinh (2r)
		e^{-2i\Phi} \delta(t-t')\\
		\avg{\hd_{\mm{in}} (t) \hd_{\mm{in}}^\dag (t')} &= \cosh^2(r) \delta(t-t').
	\end{aligned}
	\label{eq:DefInputField}
\end{equation}
Here, $r$ denotes the squeezing parameter, $\Phi$ is the angle along which the field is squeezed,
and $\delta(x)$ is the Dirac delta function.

From the standard input-output relations, the output field is
\begin{equation}
	\hd_{\mm{out}} (t) = \hd_{\mm{in}} (t) - \sqrt{\kappa} \hd (t),
	\label{eq:InOutRel}
\end{equation}
from which we can calculate the phase quadrature that can be measured during a homodyne
measurement:
\begin{equation}
	\begin{aligned}
		\hQ (t) &= i \sqrt{\kappa}\left[ \hd_{\mm{out}}^\dag (t) - \hd_{\mm{out}} (t)\right],
	\end{aligned}
	\label{eq:HomodyneSignals}
\end{equation}
which is the ideal information-containing quadrature.

As is standard, the estimator for the state of the qubit will be constructed from the time-integrated homodyne current (integration time $\tau$).  We thus define:
\begin{equation}
	\begin{aligned}
		\bar{Q}_z (\tau) &= \int_0^\tau \ud t \avg{\hQ (t)}_z,\\
		\bar{Q}^2_z (\tau) &= \int_0^\tau \ud t \int_0^\tau \ud t' \avg{\hQ (t) \hQ^\dag
		(t')}_z,
	\end{aligned}
	\label{eq:IntSignals}
\end{equation}
with $z=\pm 1 = \uparrow,\downarrow$.

Within this framework, the signal-to-noise ratio is given by
\begin{equation}
	\mm{SNR} = \frac{\abs{\bar{Q}_\uparrow - \bar{Q}_\downarrow}^2}{\bar{Q}^2_\uparrow +
	\bar{Q}^2_\downarrow}.
	\label{eq:SNR}
\end{equation}

To illustrate how a closed form expression can be obtained for Eq.~\eqref{eq:SNR}, we consider the
simple case of evaluating the average output field. We have
\begin{equation}
	\begin{aligned}
		\avg{\hd_{\mm{out}} (t)} &= -\sqrt{\kappa} \avg{\hd (t)} \\
		&=  2 i \sqrt{\kappa} \beta_z \Omega_\uR \bar{a}_0 \exp\left[-\frac{\kappa}{2} t\right] \exp\left[-i
		\beta_z \sin\left(\Omega_\uR t\right)\right] D(t),
	\end{aligned}
	\label{eq:doutAvg}
\end{equation}
with
\begin{equation}
	\begin{aligned}
		D (t) &= \int_{-\infty}^t \ud t_1 \exp\left[\frac{\kappa}{2} t_1\right]
		\exp\left[i \beta_z \sin\left(\Omega_\uR t_1\right)\right]
		\cos^2\left(\Omega_\uR t_1\right) \\
		&= \sum_{k=-\infty}^\infty J_k (\beta_z) \int_{-\infty}^t \ud t_1
		\exp\left[\frac{\kappa}{2} t_1\right] \exp\left[i k \Omega_\uR t_1\right]
		\cos^2\left(\Omega_\uR t_1\right) \\
		&= \frac{1}{2}\exp\left[\frac{\kappa}{2} t\right] \sum_{k=-\infty}^{\infty} J_k
		(\beta_z) \left( \frac{e^{i(k-2)\Omega_\uR t}}{\kappa+ 2 i (k-2)\Omega_\uR} + 2
		\frac{e^{i k \Omega_\uR t}}{\kappa + 2 i k \Omega_\uR} + \frac{e^{i(k+2)\Omega_\uR
		t}}{\kappa+ 2 i (k+2)\Omega_\uR}\right).
	\end{aligned}
	\label{eq:intd}
\end{equation}
The second equality follows from the Jacobi-Anger expansion:
\begin{equation}
	\exp[i x \sin(y)] = \sum_{k=-\infty}^\infty J_k (x) \exp[i k y].
	\label{eq:JacobiAnger}
\end{equation}
Here, $J_k (x)$ is the Bessel function of the first kind.

Given that $\left| \beta_z \right| \ll 1$ (c.f. Eq.~\eqref{eq:Bz}), we can expand Eq.~\eqref{eq:doutAvg} in powers of
$\beta_z$. Although the calculations are more elaborate, we can proceed similarly to compute the
correlators of the output field. Finally, after a lengthy but straightforward calculation, we find
the average integrated signal to be
\begin{equation}
	\abs{\bar{Q}_\uparrow (\tau) - \bar{Q}_\downarrow (\tau)}^2 = 64 \kappa^2 \chi^2 n
	\left[F_0^2 (\tau) + \beta^2 F_0 (\tau) R (\tau)\right] + \mathcal{O}\left(\beta^4\right),
	\label{eq:AvgIntDiffSignSq}
\end{equation}
with $\beta=\chi/\Omega_\uR$,
\begin{equation}
	F_0 (\tau) = \frac{\tau}{\kappa} + \frac{\sin\left(\Omega_\uR \tau\right)}{\kappa^2 +
	16\Omega_\uR^2} \left[4 \sin\left(\Omega_\uR \tau\right) +
	\frac{\kappa}{\Omega_\uR}\cos\left(\Omega_\uR \tau\right)\right],
	\label{eq:G0}
\end{equation}
and
\begin{equation}
	\begin{aligned}
		R (\tau) &= -\frac{3 \tau}{4 \kappa} + \frac{\kappa \tau}{4\left(\kappa^2 +
		16\Omega_\uR^2\right)} + \frac{\kappa \tau}{2\left(\kappa^2 + 4
		\Omega_\uR^2\right)} + \frac{\sin\left(2 \Omega_\uR \tau\right)}{4 \kappa \Omega_\uR}\\
		&\phantom{={}}
		- \frac{\sin\left(\Omega_\uR \tau\right)}{\kappa^2 +
		4\Omega_\uR^2}\left[\sin\left(\Omega_\uR \tau\right) + \frac{\kappa}{2
		\Omega_\uR}\cos\left(\Omega_\uR \tau\right)\right] \\
		&\phantom{={}}
		- \frac{\sin^2\left(\Omega_\uR \tau\right)}{\kappa^2 +
		36\Omega_\uR^2}\left[3 \cos\left(2 \Omega_\uR \tau\right) -
		\frac{\kappa}{2\Omega_\uR}\sin\left(2 \Omega_\uR \tau\right)\right] \\
		&\phantom{={}}
		-\frac{\sin\left(\Omega_\uR \tau\right)}{\kappa^2 + 16\Omega_\uR^2}\left[2
		\sin^3\left(\Omega_\uR \tau\right) + \frac{\kappa}{2\Omega_\uR}
		\cos\left(\Omega_\uR \tau\right) - \frac{\kappa}{4\Omega_\uR}\cos\left(\Omega_\uR
		\tau\right)\cos\left(2 \Omega_\uR \tau\right)  \right] \\
		&\phantom{={}}
		+ \frac{\sin\left(2 \Omega_\uR \tau\right)}{\kappa^2 + 64 \Omega_\uR^2}\left[
		\sin\left(2 \Omega_\uR \tau\right) + \frac{\kappa}{8\Omega_\uR}
		\cos\left(2 \Omega_\uR \tau\right)\right].
	\end{aligned}
	\label{eq:R}
\end{equation}

The integrated noise is
\begin{equation}
	\begin{aligned}
		\bar{Q}^2_\uparrow + \bar{Q}^2_\downarrow &= 2\left[\cosh(2r) - \cos(2\theta)
		\sinh(2r)\right] \kappa \tau + \beta^2 \sinh(2r) \cos(2\theta) \times \\
		&\phantom{={}}
		\left\{ \frac{32 \Omega_\uR^2}{\kappa^2 + 4 \Omega_\uR^2} \kappa \tau
		- \frac{16 \Omega_\uR^2 \left(5 \kappa^4 + 16 \Omega_\uR^4\right)}
		{\left(\kappa^2 + 4\Omega_\uR^2\right)^2\left(\kappa^2 +\Omega_\uR^2\right)}
		\right. \\
		&\phantom{={}}
		+ \frac{16 \kappa^2 \Omega_\uR
		\left[ \Omega_\uR \left(-7\kappa^2 +8 \Omega_\uR^2\right)
		\cos\left(2 \Omega_\uR \tau\right) + \kappa\left(\kappa^2 - 14 \Omega_\uR^2\right)
		\sin\left(2 \Omega_\uR \tau\right)\right]}{\left(\kappa^2+
		\Omega_\uR^2\right)\left(\kappa^2 + 4 \Omega_\uR^2\right)\left(\kappa^2 + 16
		\Omega_\uR^2\right)} \\
		&\phantom{={}}
		+16 e^{-\frac{\kappa \tau}{2}} \left[
		\frac{4 \Omega_\uR^2\left(2\kappa^4 +3\kappa^2 \Omega_\uR^2 +16 \Omega_\uR^4\right)}
		{\left(\kappa^2+\Omega_\uR^2\right)\left(\kappa^2 +4\Omega_\uR^2\right)\left(\kappa^2 + 16\Omega_\uR^2\right)}
		\right.\\
		&\phantom{=+16 e^{-\frac{\kappa \tau}{2}} [}
		\left. \left.
		+ \frac{2 \kappa^2 \Omega_\uR\left(\kappa^2 -2 \Omega_\uR^2\right)\left[2 \Omega_\uR
		\cos\left(\Omega_\uR \tau\right) + \kappa \sin\left(2 \Omega_\uR \tau\right)\right]}
		{\left(\kappa^2 + 4\Omega_\uR^2\right)^2\left(\kappa^2 + \Omega_\uR^2\right)} \right]
		\right\}.
	\end{aligned}
	\label{eq:VarQ}
\end{equation}

Using Eqs.~\eqref{eq:AvgIntDiffSignSq} and \eqref{eq:VarQ}, we can evaluate the improvement of the
SNR in the long-time limit, $\kappa \tau \gg 1$. We find
\begin{equation}
	\frac{\mm{SNR} (r)}{\mm{SNR} (r=0)} \simeq
	\frac{e^{2r}}{1+2\beta^2\left(1-\frac{\gamma^2}{4}+\frac{\gamma^4}{16}\right)\left(e^{4r}-1\right)},
	\label{eq:SNRRatioLongTime}
\end{equation}
which is a non-monotonic function of the input squeezing parameter $r$. The maximum of Eq.~\eqref{eq:SNRRatioLongTime} (corresponding to the optimal amount of input squeezing) is given by
\begin{equation}
	e^{2r_{\mm{opt}}}= \sqrt{\frac{8 -\beta^2\left(16 - 4\gamma^2 +
	\gamma^4\right)}{\beta^2\left(16 - 4\gamma^2 + \gamma^4\right)}}.
	\label{eq:Optr}
\end{equation}
This shows that the improvement (i.e. the factor $e^{2r}$) of the $\mm{SNR}^2$ when using the optimal $r$ scales like
$1/\beta$ in the long time limit.

By replacing $r$ in Eq.~\eqref{eq:SNRRatioLongTime} with $r_{\mm{opt}}$ as given by
Eq.~\eqref{eq:Optr}, we find by how much the $\mm{SNR}^2$ can be improved. We get
\begin{equation}
	\begin{aligned}
		\frac{\mm{SNR}^2 (r_{\mm{opt}})}{\mm{SNR}^2 (r=0)} &\simeq
		\frac{1}{2\sqrt{2}\beta\sqrt{1-2\beta^2}} + \frac{-1 + 8\beta^2
		\gamma^4}{256\sqrt{2}\beta \left( 1 -2 \beta^2\right)^\frac{5}{2}} +
		\frac{\left(1 - 4\beta^2\right)\gamma^2}{16\sqrt{2}\beta\left( 1 -2
		\beta^2\right)^\frac{3}{2}} \\
		& \simeq \frac{1+\beta^2}{2\sqrt{2}\beta} + \frac{\gamma^2}{16\sqrt{2}\beta}.
	\end{aligned}
	\label{eq:SNRSquareImprov}
\end{equation}

Using the experimental values of the different parameters, we find $e^{2r_{\mm{opt}}} = 15.9\,\mm{dB}$ and $\mm{SNR}^2 (r_{\mm{opt}})/\mm{SNR}^2 (r=0) \simeq 19.4$.  We stress that in general, even if the dispersive coupling is optimally strong (i.e. $\chi \simeq \kappa$), for $\beta \ll 1$ the SNR enhancement from squeezing can be extremely large.  This is in marked contrast from injected squeezing in a standard dispersive setup, where negligible enhancement is possible when $\chi \simeq \kappa$ \cite{Barzanjeh2014DispersiveAmplifiers,Didier2015HeisenbergLight}.

%%%%%%%%%%%%%%% THEORY SECTION 3 %%%%%%%%%%%%%%%%

\section{Qubit decay induced by the counter-rotating terms}

For perfect synthetic longitudinal coupling, $\sz$ commutes with the full Hamiltonian, and there is no qubit decay or excitation induced by measurement. However, due to finite $\Omega_R$, non-RWA terms in the full Hamiltonian (i.e. $\hat{A}$ and $\hat{B}$ in Eq.~\eqref{eq:HBAE}) can cause non-QND effects. In particular, the terms proportional to $\sy$ can cause qubit decay or excitation, and in this section we will assess these effects.

We will examine both the situation of a broadband squeezed vacuum incident on the cavity, and the more experimentally relevant situation of an incident squeezed state with finite bandwidth $\gamma$. Simulation parameters are chosen to match the experimental values:
\begin{table}[h!]
\begin{centering}
  \begin{tabular}{ | l | c | }
    \hline
    $\kappa/2\pi$ & 5.9 MHz \\ \hline
    $\chi/2\pi$ & 0.73 MHz  \\ \hline
    $\Omega/2\pi$ & 40 MHz  \\ \hline
    $\bar{a}_0$ & $0.35$  \\ \hline
    $\kappa_{\rm SQZ}/2\pi$ & 26 MHz  \\
    \hline
  \end{tabular}
\end{centering}
    \caption{Table of simulation parameter values.}\label{tab:Pars}
\end{table}

For an infinite bandwidth (broadband) input squeezed state with photon number $N_s = \sinh^2(r)$, the master equation for the evolution of the qubit-cavity system is \cite{QuantumNoise}
\begin{align}
  \dot{\rho} = -i\left[\hH_{\rm R},\rho\right] + \kappa\mathcal{D}\left[\sqrt{N_s + 1}\hat{d} - e^{i2\Phi}\sqrt{N_s}\hat{d}^\dagger\right]\rho, \label{eqn:MEBroad}
\end{align}
where $\hH_{\rm R}$ is the Hamiltonian of Eq.~\eqref{eq:HBAE}, and the angle $\Phi$ defines which input quadrature is squeezed. We will consider $\hat{X} = (\hd + \hd^\dagger)/\sqrt{2}$ quadrature squeezing ($\Phi = 0$) as this corresponds to an enhancement in the measurement rate.

To describe finite bandwidth input squeezing, we model the source of the squeezing as a degenerate parametric amplifier (DPA) whose output is the input of our qubit-cavity system. This can be described by the cascaded master equation
\begin{align}
    \dot{\rho} = -i\left[\hH_{\rm cas},\rho\right] + \mathcal{D}\left[\sqrt{\kappa_{\rm SQZ}}\hat{b} + \sqrt{\kappa}\hat{d}\right]\rho, \label{eqn:MECas}
\end{align}
where $\hat{b}$ is the lowering operator for the DPA, and $\kappa_{\rm SQZ}$ is the decay rate from the DPA's output port (which sets the corresponding bandwidth of the output squeezing). The Hamiltonian $\hH_{\rm cas}$ is given by
\begin{align}
  \hH_{\rm cas} = \hH_{\rm R}+ i\lambda\left(e^{i2\Phi}\hat{b}^{\dagger 2} -e^{-i2\Phi}\hat{b}^2\right) + i\frac{\sqrt{\kappa_{\rm SQZ}\kappa}}{2}\left(\hat{d}\hat{b}^\dagger - \hat{d}^\dagger\hat{b}\right),
\end{align}
where as before $\Phi$ is the squeezing angle, and $\lambda$ is the parametric drive strength that sets the output squeezed photon number.

To ensure a fair comparison we keep the intracavity photon number at steady state the same for both the broadband and finite bandwidth simulations. The steady state intracavity photon number for the cascaded simulation is given by
\begin{align}
  N_s = \frac{2\kappa_{\rm SQZ}\lambda^2(2\kappa_{\rm SQZ} + \kappa)}{\left[\left(\frac{\kappa_{\rm SQZ}}{2}\right)^2-\left(2\lambda\right)^2\right]\left[\left(\frac{\kappa_{\rm SQZ}}{2}\right)^2-\left(2\lambda\right)^2 + \frac{\kappa_{\rm SQZ}\kappa}{2} + \left(\frac{\kappa}{2}\right)^2\right]},
\end{align}
from which we can calculate the value of $\lambda$ for a given $N_s$.

\begin{figure}
\includegraphics[width=0.9\textwidth]{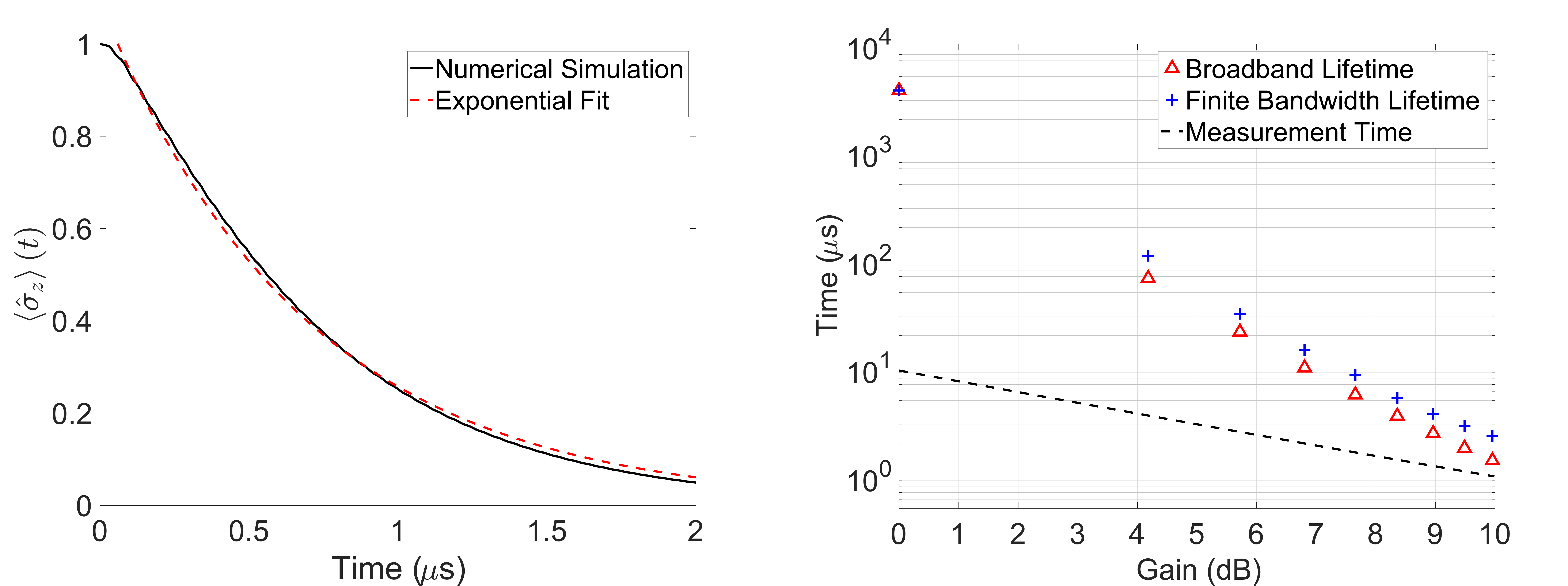}
  \caption{(Left Panel) Short time evolution of the expectation value of $\sz$ (black curve) for approximately 10 dB of input squeezing, which can be well fit by an exponential (red dashed line). (Right Panel) Effective qubit lifetime ($T_{\rm eff}$) as a function of the input squeezing power measured in terms of the amplifier gain of the DPA (in units of dB), for both broadband (red triangles) and finite bandwidth (blue crosses) input squeezing. Finite bandwidth squeezing improves the qubit lifetime as compared to broadband squeezing, and in both cases, the effective qubit lifetime is longer than the time taken to reach a measurement fidelity of $99.9\%$, shown by the dashed black line (calculated using Eqs.~\eqref{eq:doutAvg} and \eqref{eq:VarQ}).\label{fig:theoryPlots}}
\end{figure}

We numerically simulate Eqs.~\eqref{eqn:MEBroad} and \eqref{eqn:MECas}, and calculate $\left<\hat{\sigma}^{\rm R}_z\right>(t) = {\rm Tr}\left[\rho(t)\mathbb{I}_{\rm c}\otimes\hat{\sigma}^{\rm R}_z\right]$. We consider the case where the squeezing is turned on at $t = 0$, such that we start with the cavity initially in vacuum and the qubit in its excited state, i.e. $\ket{0}_{\rm c}\ket{e}_{\rm q}$ is the initial state in the Rabi frame. The evolution of the qubit state due to the non-QND coupling is not purely incoherent, especially at larger $N_s$, as the cavity effectively acts as a complex non-Markovian bath. However, the short time decay of the qubit is well fit by an exponential
\begin{align}
\left<\hat{\sigma}^{\rm R}_z\right>(t) = e^{-\frac{2t}{T_{\rm eff}}},
\end{align}
as shown in the left panel of \FigSThree \ for the maximum simulated DPA gain of 10 dB. We use this fit to extract the effective qubit lifetime, $T_{\rm eff}$, for a given $N_s$ of broadband or finite bandwidth squeezing. This effective lifetime is shown in the right panel of \FigSThree, and as expected, the lifetime increases for finite bandwidth squeezing. We emphasize that this effective lifetime describes only the short time decay of the qubit, and quantifies how quickly non-QND terms in the full Hamiltonian (c.f. Eq.~\eqref{eq:HBAE}) scramble the initial qubit state information, such that measurement fails.

We compare the effective qubit lifetime to the measurement time necessary to reach a readout fidelity of $99.9\%$, which can be calculated using the full expression for the SNR derived from Eqs.~\eqref{eq:doutAvg} and \eqref{eq:VarQ}. As the measurement time is very short for large $N_s$, we must calculate the measurement time using the full expression for the SNR, rather than the long time expression for the measurement rate found in the main text. As can be seen in the right panel of \FigSThree, the effective qubit lifetime is longer than the measurement time for both broadband and finite bandwidth squeezing.

If we were to instead consider $\hat{Y} = i(\hd^\dagger - \hd)/\sqrt{2}$ quadrature squeezing ($\Phi = \pi/2$), then the effective qubit lifetime would decrease. This is due to the fact that for $\hat{Y}$ quadrature squeezing, fluctuations in the $\hat{X}$ quadrature are amplified, which enhances the effect of the counter-rotating term $\chi a_0\sin(2\Omega_Rt)\hat{\sigma}^{\rm R}_y\left(\hat{d} + \hat{d}^\dagger\right)$. This shows that the effective qubit lifetime depends not only on the number of squeezed photons, but also the phase of the input squeezed state. Conveniently, the longest qubit lifetime is found for the mode of operation that has the shortest measurement time, i.e. $\hat{X}$ quadrature squeezing.

%
% ****** End of file apssamp.tex ******

%%%%%%%%%% REFERENCES %%%%%%%%%%
\bibliography{Mendeley,customEntries}

\end{document}